\documentclass[superscriptaddress,twocolumn,secnumarabic,
amssymb,amsmath,nobibnotes,aps,prd,nofootinbib]{revtex4}%
\usepackage{graphicx}
\usepackage{epsf}
\usepackage{bm}
\usepackage{amsmath}
\usepackage{amsfonts}
\usepackage{amssymb}
\usepackage{epstopdf}
\usepackage{natbib}
\usepackage{color}%
\providecommand{\U}[1]{\protect\rule{.1in}{.1in}}
\newcommand{\be}{\begin{equation}}
\newcommand{\ee}{\end{equation}}

\newcommand{\mincir}{\raise
-3.truept\hbox{\rlap{\hbox{$\sim$}}\raise4.truept\hbox{$<$}\ }}
\newcommand{\magcir}{\raise
-3.truept\hbox{\rlap{\hbox{$\sim$}}\raise4.truept\hbox{$>$}\ }}

\usepackage{color}
\usepackage{hyperref}

\begin{document}

\title{Generalized Emergent Dark Energy Model and the Hubble Constant Tension}

\author{Weiqiang Yang}
\email{d11102004@163.com}
\affiliation{Department of Physics, Liaoning Normal University, Dalian, 116029, P. R. China}

\author{Eleonora Di Valentino}
\email{eleonora.di-valentino@durham.ac.uk}
\affiliation{Institute for Particle Physics Phenomenology, Department of Physics, Durham University, Durham DH1 3LE, UK}

\author{Supriya Pan}
\email{supriya.maths@presiuniv.ac.in}
\affiliation{Department of Mathematics, Presidency University, 86/1 College Street, 
Kolkata 700073, India}

\author{Arman Shafieloo}
\email{shafieloo@kasi.re.kr}
\affiliation{Korea Astronomy and Space Science Institute, Daejeon 34055, Korea}
\affiliation{University of Science and Technology, Yuseong-gu 217 Gajeong-ro, Daejeon 34113, Korea} 

\author{Xiaolei Li}
\email{lixiaolei@hebtu.edu.cn}
\affiliation{College of Physics, Hebei Normal University, Shijiazhuang 050024, China}

\begin{abstract}
We investigate a generalized form of the phenomenologically emergent dark energy model, known as generalized emergent dark energy (GEDE), introduced by Li and Shafieloo  [Astrophys. J. {\bf 902}, 58 (2020)] in light of a series of cosmological probes and considering the evolution of the model at the level of linear perturbations. This model introduces a free parameter $\Delta$ that can discriminate between the $\Lambda$CDM  (corresponds to $\Delta=0$) or the phenomenologically emergent dark energy (PEDE) (corresponds to $\Delta=1$) models, allowing us to determine which  model is preferred most by the fit of the observational datasets. We find evidence in favor of the GEDE model for Planck alone and in combination with R19, while the Bayesian model comparison is inconclusive when Supernovae Type Ia or BAO data are included. In particular, we find that $\Lambda$CDM model is disfavored at more than $2\sigma$ CL for most of the observational datasets considered in this work and PEDE is in agreement with Planck 2018+BAO+R19 combination within $1\sigma$ CL. 
\end{abstract}

\maketitle
\section{Introduction}

With the growing sensitivity in experimental data, cosmology is becoming richer than it was several years back. The observational probes suggest that the main ingredients of the universe are two dark fluids, one is pressureless dark matter (DM) or cold dark matter (CDM) and one is some hypothetical dark energy (DE). This DE fluid is solely responsible for the observed accelerating expansion of the universe. This observational description can be described in the framework of the $\Lambda$CDM cosmology where the $\Lambda$ term $-$ independent of the cosmic evolution, acts as the so-called DE fluid. The $\Lambda$CDM model has received the greatest focus for its ability to fit most of the observational data while it cannot explain the actual physics of DM and DE. If we avoid the physics behind DM and DE within the $\Lambda$CDM picture, still it cannot explain some very serious issues. The tensions in the Hubble constant $H_0$~\cite{Riess:2019cxk,Riess:2020fzl,Aghanim:2018eyx,Verde:2019ivm,DiValentino:2020vnx} and in the amplitude of the growth of structure $\sigma_8$ (alternatively one can also measure the parameter $S_8 \equiv \sigma_8 \sqrt{\Omega_{m0}/0.3}$) are the signals of the limitations of the $\Lambda$CDM cosmology (see Refs.~\cite{DiValentino:2020zio,DiValentino:2020vvd} for a recent overview, and references therein). Therefore, the search for some alternative descriptions to the $\Lambda$CDM cosmology having explanations to the aforementioned tensions in these cosmological parameters, has been the heart of modern cosmology now. Usually, the alleviation or solution of the tensions is achieved by extending the parameter space of the six parameter $\Lambda$CDM model. This extension is performed in many ways. 

For example, early modifications of the expansion history are promising because such modifications can increase the $H_0$ value and give a lower sound horizon $r_{\mathrm{drag}}$ at the drag epoch~\cite{Knox:2019rjx,Evslin:2017qdn}, in agreement with the Baryon Acoustic Oscillations (BAO) data. 
In this category we find the Early Dark Energy models~\cite{Pettorino:2013ia,Poulin:2018cxd,Karwal:2016vyq,Sakstein:2019fmf,Niedermann:2019olb,Akarsu:2019hmw,Gogoi:2020qif,Smith:2019ihp,Hill:2020osr,Murgia:2020ryi,Lucca:2020fgp,Haridasu:2020pms,Smith:2020rxx,DAmico:2020ods,Chudaykin:2020igl,Kaloper:2019lpl,Chudaykin:2020acu,Berghaus:2019cls,Alexander:2019rsc,Agrawal:2019lmo,Niedermann:2020dwg,Ye:2020btb,Lin:2019qug,Lin:2020jcb,Yin:2020dwl,Braglia:2020bym}),
and extra relativistic species at recombination~\cite{Vagnozzi:2019ezj,Seto:2021xua,Carneiro:2018xwq,Gelmini:2019deq,Gelmini:2020ekg,Barenboim:2016lxv,DEramo:2018vss,Pandey:2019plg,Xiao:2019ccl,Nygaard:2020sow,Blinov:2020uvz,Binder:2017lkj,Hryczuk:2020jhi,Vattis:2019efj,Haridasu:2020xaa,Clark:2020miy,Gu:2020ozv,Alcaniz:2019kah,Choi:2019jck,Escudero:2019gzq,Anchordoqui:2020znj,Desai:2019pvs,Choi:2020tqp,DiValentino:2017oaw,Escudero:2019gvw,Arias-Aragon:2020qip,Blinov:2020hmc,Buen-Abad:2018mas,Gonzalez:2020fdy,Flambaum:2019cih,Anchordoqui:2019yzc,Flores:2020drq,Artymowski:2020zwy}, but they cannot solve the tension with R19 below $3\sigma$~\cite{Arendse:2019hev,Lin:2021sfs}. Features in the form of the primordial power spectrum have been proposed as another possible early universe solution for the Hubble tension problem~\cite{Hazra:2018opk,Keeley:2020rmo} but these models still need to be tested against a combination of various observations.   
One can also modify the physics at late time to explain the Hubble tension issue. 
While the late time modifications of the expansion history can solve the Hubble tension with R19 within $1\sigma$ CL, they do not modify the sound horizon, resulting in tension with the BAO data. In this category, we have Phantom Dark Energy models~\cite{Aghanim:2018eyx,Vagnozzi:2019ezj,Martinelli:2019krf,Alestas:2020mvb,DiValentino:2020vnx,Haridasu:2020pms,Menci:2020ybl,Kitazawa:2020qdx,DiValentino:2019dzu,Yang:2021flj,Yang:2018qmz,DiValentino:2020naf}), Emergent Dark Energy~\cite{Li:2019yem,Pan:2019hac,Rezaei:2020mrj,Liu:2020vgn,Li:2020ybr,Hernandez-Almada:2020uyr,Yang:2021egn}, and additional Dark Energy proposals~\cite{Banihashemi:2018oxo,Miao:2018zpw,Banihashemi:2018has,Li:2019san,Yang:2020zuk,Banihashemi:2020wtb,Camarena:2021jlr,Benevento:2020fev,Alestas:2020zol,Sola:2017znb,Keeley:2019esp,Dutta:2018vmq,daSilva:2020mvk,Yang:2019qza,Guo:2018ans,daSilva:2020bdc,Agrawal:2019dlm,Colgain:2018wgk,Colgain:2019joh,Banerjee:2020xcn,DiValentino:2019exe,Adler:2019fnp}. Other promising Hubble tension solutions are models considering a non-gravitational interaction between DM and DE, widely known as the Interacting Dark Energy (IDE) models~\cite{Kumar:2017dnp,DiValentino:2019ffd,Kumar:2019wfs,DiValentino:2020leo,Lucca:2020zjb,DiValentino:2020vnx,Yang:2019uog,Martinelli:2019dau,Gomez-Valent:2020mqn,DiValentino:2017iww,Yang:2020uga,Kumar:2016zpg,Yang:2018euj,Pan:2020bur,Yao:2020hkw,Yao:2020pji,Pan:2019gop,Yang:2019uzo,Pan:2019jqh,Yang:2018ubt,Amirhashchi:2020qep,Gao:2021xnk}. Concerning the tension in the Hubble constant, we refer to an exhaustive review that appeared recently \cite{DiValentino:2021izs}.  

Effectively, in almost every work, an extension of the parameter space (compared to the minimal $\Lambda$CDM scenario) is needed to alleviate or solve this tension.  However, both  approaches, that means the early and late modifications of the expansion history of the universe, as described above have some shortcomings. Although a model with extended parameter space compared to the minimal $\Lambda$CDM model, can address the $H_0$ tension, at the same time, due to the presence of additional free parameters, the fitting of the model gets worse compared to the excellent fitting that we witnessed in the  $\Lambda$CDM model. A natural question in this context is therefore raised: whether one could find some cosmological models having exactly six free parameters as in the $\Lambda$CDM model, but at the same time the model can alleviate or solve the $H_0$ tension.
There are only a very few works where the $H_0$ tension can be solved without adding any extra parameter from outside. One is the vacuum metamorphosis model~\cite{DiValentino:2017rcr,DiValentino:2020kha} and the other one is the PEDE scenario, see for instance Refs.~\cite{Li:2019yem,Pan:2019hac}. Both models are excellent to reconcile the $H_0$ tension but the problem with the sound horizon still remains. Nevertheless, both  models offer excellent fits to some of the observational datasets, but not for all though. So, the cosmological models with six free parameters can be considered to be the potential contenders of other cosmological models.

Having such appealing possibilities associated with these six-parameter based cosmological models, in the present article we focus only on an extended version of the PEDE model known as the Generalized Emergent Dark Energy (GEDE) model~\cite{Li:2020ybr} (also see \cite{Hernandez-Almada:2020uyr}). The GEDE model contains an extra free parameter $\Delta$ which recovers both $\Lambda$CDM ($\Delta = 0$) and PEDE model ($\Delta = 1$) as special cases. Certainly, this GEDE model allows a wider cosmological picture between the $\Lambda$CDM model and the PEDE model and any departure from $\Delta  =0$ will strengthen the viability of the new scenario. We investigated this scenario including the evolution of the model at the level of linear perturbations in order to examine whether this extended version of PEDE can equally release the $H_0$ tension when almost all cosmological probes are taken into account. We quickly note here that GEDE 
can alleviate the $H_0$ tension and offer an excellent fit to most of the observational data employed in this work, when we compare with the $\Lambda$CDM model.

The article has been organized in the following way. In section \ref{sec-GEDE} we review the GEDE model. After that in section \ref{sec-data} we describe the observational data and the statistical methodology for the fitting of the model.  In the next section \ref{sec-results} we discuss the observational constraints and the $H_0$ problem in detail. Finally, in section \ref{sec-conclu} we summarize our results focusing on the main ingredients of the model.

\section{Reintroducing Generalized Emergent Dark energy}
\label{sec-GEDE}

We consider a spatially flat Friedmann-Lema\^{i}tre-Robertson-Walker (FLRW) Universe where radiation, pressureless matter (baryons plus cold dark matter) and a DE fluid are present. All are assumed to be barotropic in nature and none of them are interacting with each other. 
In such a background, one can readily write down the Hubble expansion as 

\begin{eqnarray}
H^2 (a) = H_0^2 \Bigl[\Omega_{r0}a^{-4}+ \Omega_{m0} a^{-3}+ \widetilde{\Omega}_{\rm{DE}}(a)\Bigr]
\end{eqnarray}
where $a$ is the scale factor of the FLRW Universe which is related to the redshift $z$ as $1+z = a_0/a$ ($a_0$ being the present value of the scale factor which we set to unity); 
$\Omega_{r0} = \rho_{r0}/\rho_{\rm crit,0}$, and $\Omega_{m0} = \rho_{m0}/\rho_{\rm crit,0}$ are, respectively, the present values of the density parameters for radiation and pressureless matter and $\widetilde{\Omega}_{\rm{DE}}(a) = \rho_{\rm DE} (a)/\rho_{\rm crit,0}$ is the energy density of the DE fluid with respect to the critical energy density at present, namely $\rho_{\rm {crit},0}(a)=3H_0^2(a)/8\pi G$  and $ \widetilde{\Omega}_{\rm{DE},0}= (1-\Omega_{m0}-\Omega_{r0})$. One can easily notice that $\widetilde{\Omega}_{\rm{DE}}(a)$ is related to the density parameter for DE ($\Omega_{\rm DE} (a)$) as follows: 

\begin{align}
   \widetilde{\Omega}_{\rm{DE}}(a) & \,  =\,\frac{\rho_{\rm{DE}}(a)}{\rho_{\rm {crit,0}}}\,=\, \frac{\rho_{\rm{DE}}(a)}{\rho_{\rm {crit}}(a)}\times \frac{\rho_{\rm {crit}}(a)}{\rho_{\rm {crit,0}}}  \\ 
  & \, =\,{\Omega}_{\rm{DE}}(a)\times \frac{H^2(a)}{H_0^2}
\end{align}
Therefore, once the evolution of $\widetilde{\Omega}_{\rm{DE}}(a)$ is known, the dynamics of the Universe can in principle be determined either analytically or numerically. In the present work we consider that $\widetilde{\Omega}_{\rm{DE}}(a)$ evolves as \cite{Li:2020ybr}

\begin{equation} \label{Omega-DE}
\widetilde{\Omega}_{\rm{DE}}(z)\,=\, \Omega_{\rm{DE,0}}\frac{ 1 - {\rm{tanh}}\left(\Delta \times {\rm{log}}_{10}(\frac{1+z}{1+z_t}) \right) }{{1+ {\rm{tanh}}\left(\Delta \times {\rm{log}}_{10}({1+z_t}) \right)}}
\end{equation}
where $\Delta$ is the only free parameter which can take any real values. The symbol $z_t$ denotes the epoch where the matter energy density and the DE density are equal, that means $\rho_m (z_t) = \rho_{\rm DE} (z_t) \implies \Omega_{m0} (1+z_t)^3 = \widetilde{\Omega}_{\rm{DE}}(z_t)$, and hence, $z_t$ is a derived parameter, not a free parameter.   

As  we have a non-interacting scenario, therefore, we use the conservation equation for DE, namely, $\dot{\rho}_{\rm DE} + 3 H (1+w_{\rm DE} (z)) \rho_{\rm DE} = 0$, where $w_{\rm DE} (z)$ is the barotropic equation of state for DE, can be derived as

\begin{equation}\label{eos-DE}
    w_{\rm DE}(z) \,=\,
    \frac{1}{3} \frac{d\,{\rm{ln}}\, {\widetilde{\Omega}_{\rm{DE}}}}{d z} (1+z)-1
\end{equation}
which for the present model (\ref{Omega-DE}), turns out to be  
\begin{equation}\label{eos--DE-explicit}
    w_{\rm DE}(z)=-1 -\frac{\Delta}{3 {\rm{ln}}\, 10} \times\left({1+{\rm{tanh}}\left(\Delta \times {\rm{log}}_{10}(\frac{1+z}{1+z_t}) \right) }\right). 
\end{equation}

Thus, the cosmological scenario, parametrized by $\tilde{\Omega}_{\rm DE}$ as shown in eqn. (\ref{Omega-DE}), is boiled down to a parametrized dynamical DE equation of state characterized by eqn. (\ref{eos-DE}) or (\ref{eos--DE-explicit}). 
With the expression for $w_{\rm DE}$, the analysis of the model at the perturbation level becomes easy and straightforward. Following \cite{Ma:1995ey}, one could derive the velocity and density perturbations for the DE fluid characterized by the  dynamical equation-of-state parameter in eqn. (\ref{eos--DE-explicit}).

\begin{figure}
    \centering
    \includegraphics[width=0.45\textwidth]{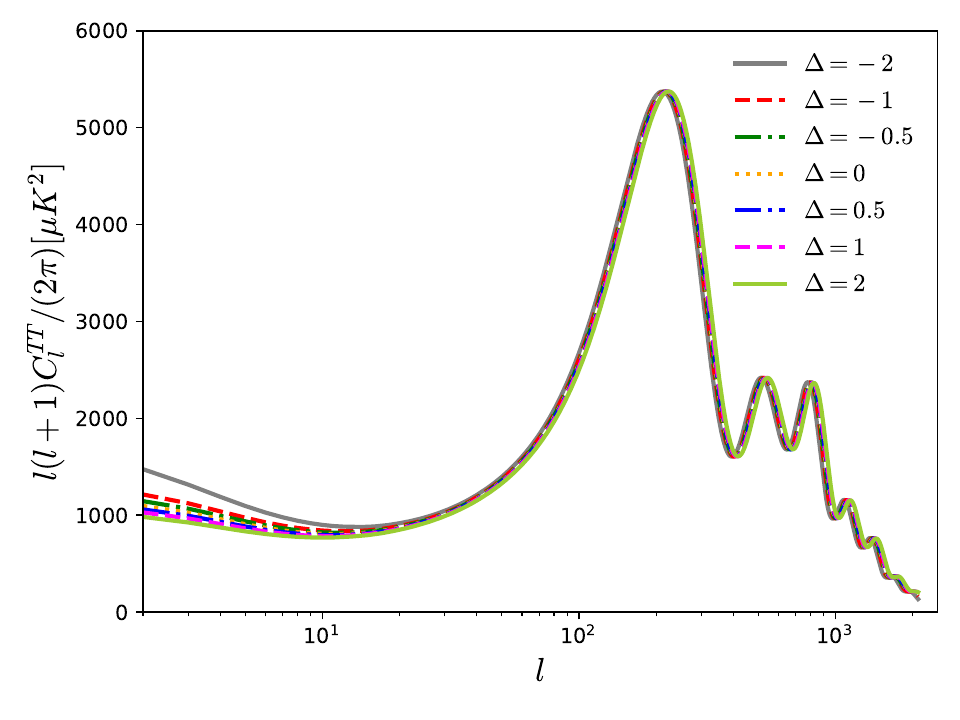}
    \includegraphics[width=0.45\textwidth]{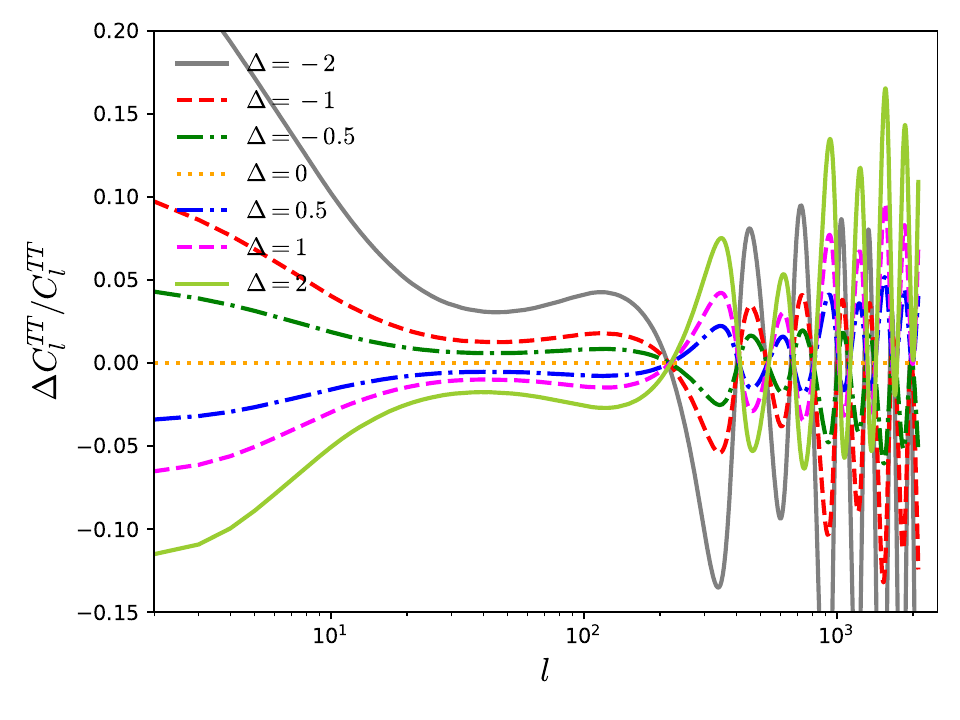}
    \caption{We show how the parameter $\Delta$ affects the temperature power spectrum (top panel) and the variation we have with respect to the $\Lambda$CDM model corresponding to $\Delta=0$ (bottom panel).}
    \label{fig:cmb-temp}
\end{figure}

\begin{figure}
    \centering
    \includegraphics[width=0.45\textwidth]{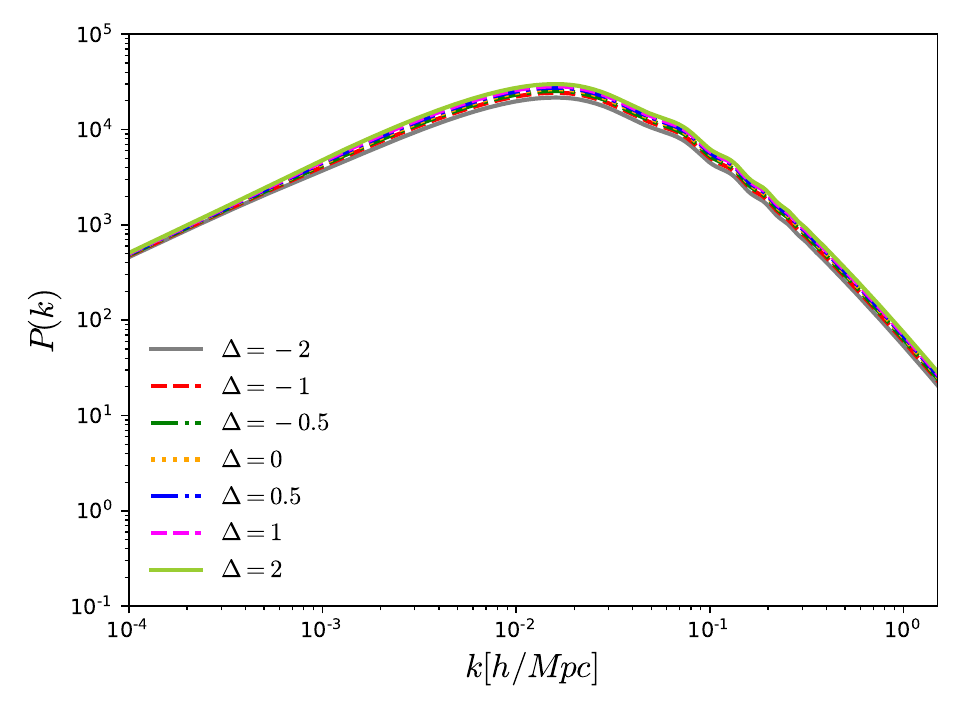}
    \includegraphics[width=0.45\textwidth]{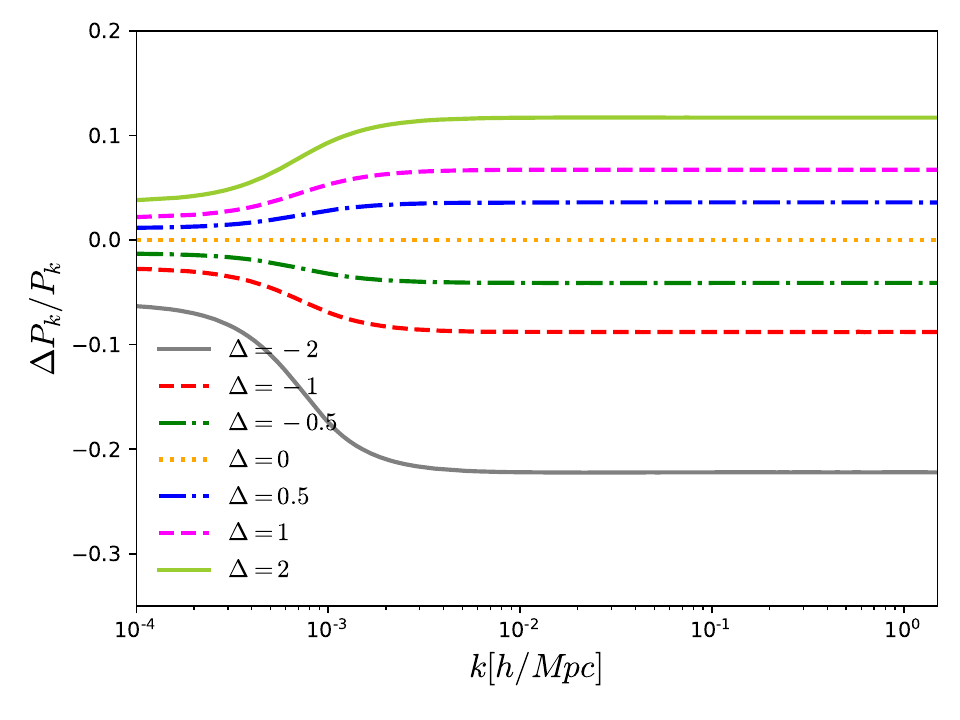}
    \caption{We show how the parameter $\Delta$ affects the matter power spectrum (top panel) and the variation we have with respect to the $\Lambda$CDM model corresponding to $\Delta=0$ (bottom panel).}
    \label{fig:matter-power}
\end{figure}

In Figs.~\ref{fig:cmb-temp} and~\ref{fig:matter-power} we show how the parameter $\Delta$ affects the temperature and the matter power spectra respectively, including the variation we have with respect to the $\Lambda$CDM model corresponding to $\Delta=0$. We can see that, increasing the value of $\Delta$ from $-2$ to $2$, we have a suppression in the amplitude of the lower multipoles and a shift of the peaks in the damping tail toward larger multipoles in the temperature power spectrum, while we have an increase of the amplitude in the matter 
power spectrum. 

\begin{figure}
    \centering
    \includegraphics[width=0.45\textwidth]{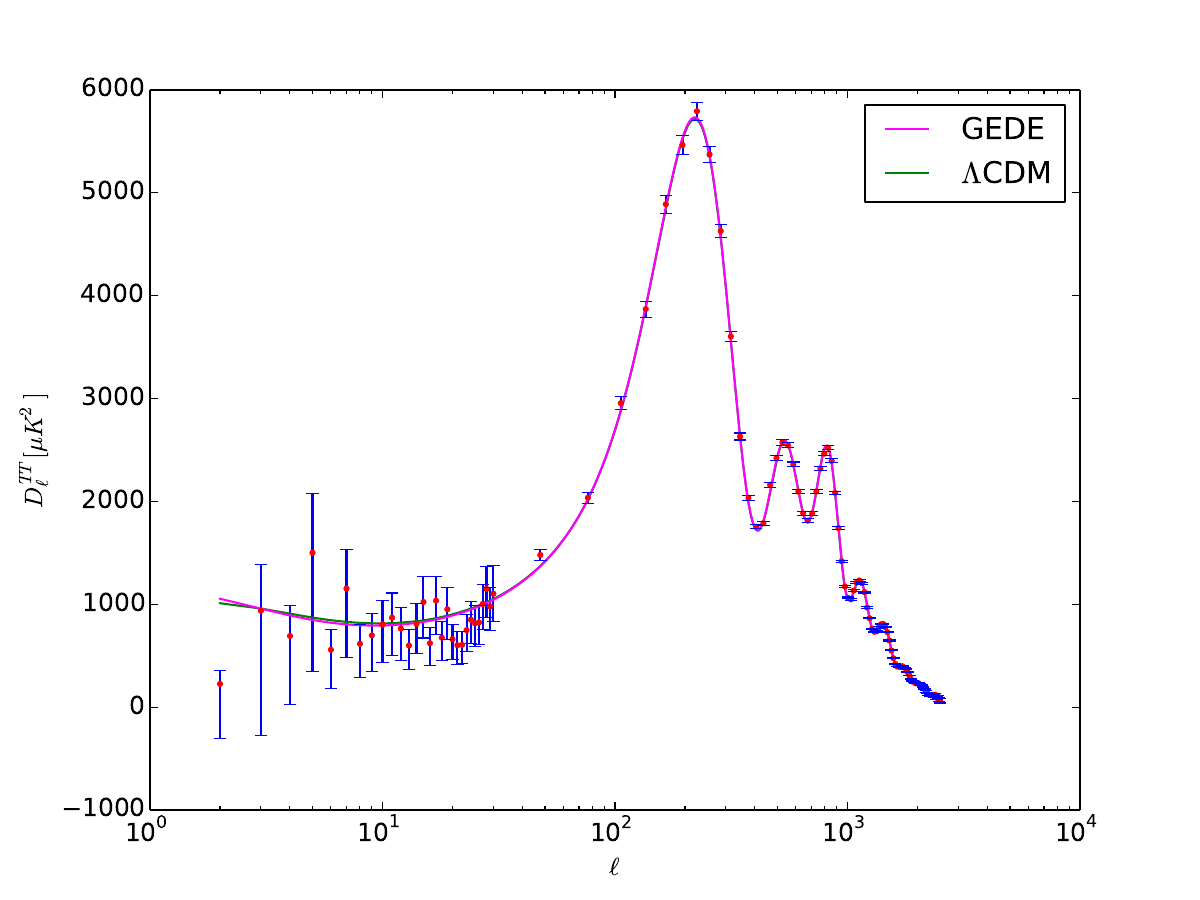}
    \caption{We display the temperature power spectrum for the cosmological scenarios, namely GEDE (solid magenta curve) and the $\Lambda$CDM scenario (green curve) obtained using the best-fit values of all the free parameters of those two scenarios from Planck 2018 alone over the Planck temperature power spectrum data points (blue vertical bars with red dot). While the two curves are indistinguishable at high multipoles, the amplitude at lower multipoles is lightly suppressed in GEDE, producing a better agreement with the data. }
    \label{fig:cmb-temp-bf}
\end{figure}

Finally, in Fig.~\ref{fig:cmb-temp-bf} we show the best fit curve  \footnote{We note that the observational constraints on GEDE are shown in Section \ref{sec-results}, as well as the best fit values of the parameters used to compute Fig. \ref{fig:cmb-temp-bf}.}
representing the CMB temperature power spectrum obtained from Planck 2018 for the GEDE model (in magenta) and the $\Lambda$CDM model (in green), together with the Planck temperature power spectrum data points. We notice that while the two curves are indistinguishable at high multipoles, the amplitude at lower multipoles is slightly suppressed in GEDE, producing a better agreement with the data.  

\section{Observational data}
\label{sec-data}

Here we describe the main observational datasets and the statistical procedure used to extract the observational constraints out of the present cosmological scenario parametrized by the DE density with respect to the critical energy density. In what follows we enlist the observational datasets: 
 
\begin{itemize}

\item CMB: We use the latest cosmic microwave background (CMB) observations from Planck 2018 legacy release, i.e. {\it plikTTTEEE+lowl+lowE}~\cite{Aghanim:2018eyx,Aghanim:2019ame}.

\item BAO: We use various measurements of the Baryon acoustic oscillation (BAO) distance measurements from 6dFGS~\cite{Beutler:2011hx}, SDSS MGS~\cite{Ross:2014qpa}, and BOSS DR12~\cite{Alam:2016hwk}.

\item $H_0$ from HST:  We consider the Hubble constant estimation from the Hubble Space Telescope measurements, using Cepheids as geometric distance calibrators, which yields to the gaussian prior $H_0 = 74.03 \pm 1.42$ km/s/Mpc at $68\%$ CL~\cite{Riess:2019cxk} (R19). 

\item Supernovae Type Ia data: We also include the latest compilation from the Pantheon catalog~\cite{Scolnic:2017caz}, or alternatively, the Joint Light-curve Analysis (JLA)~\cite{Betoule:2014frx} sample.

\end{itemize}

To extract the constraints of the GEDE scenario, we have modified the well known \texttt{CosmoMC} package~\cite{Lewis:2002ah,Lewis:1999bs}. The \texttt{CosmoMC} package is freely available  \url{http://cosmologist.info/cosmomc/} and this has a convergence diagnostic based on the Gelman-Rubin criterion~\cite{Gelman:1992zz}. Additionally, this also supports the Planck 2018 likelihood~\cite{Aghanim:2019ame}. In Table~\ref{priors}, we have shown the flat priors imposed on the free parameters of the present cosmological scenario during the statistical analysis.

\begin{table}
\begin{center}
\begin{tabular}{|c|c|c|c|c|}
\hline
Parameter                    & Prior \\
\hline
$\Omega_{\rm b} h^2$         & $[0.005,0.1]$ \\
$\Omega_{\rm c} h^2$         & $[0.001,0.99]$ \\
$100\theta_{MC}$             & $[0.5,10]$ \\
$\tau$                       & $[0.01,0.8]$ \\
$n_\mathrm{S}$               & $[0.7,1.3]$ \\
$\log[10^{10}A_{s}]$         & $[1.7, 5.0]$ \\
$\Delta$                     & $[-10,10]$  \\
\hline 
\end{tabular}
\end{center}
\caption{Flat priors on the free cosmological parameters of the GEDE scenario.  }
\label{priors}
\end{table}
\begingroup                                                                                                                     
\squeezetable                                                                                                                   
\begin{center}                                                                                                                  
\begin{table*}                                                                                                                   
\begin{tabular}{cccccccccc}                                                                                                            
\hline\hline                                                                                                                    
Parameters & Planck 2018 & Planck 2018+BAO & Planck 2018+R19 & Planck 2018+BAO+R19 \\ \hline
$\Omega_c h^2$ & $    0.1198_{-    0.0013-    0.0027}^{+    0.0013+    0.0026}$ & $    0.1198_{-    0.0012-    0.0025}^{+    0.0012+    0.0027}$  & $    0.1201_{-    0.0013-    0.0027}^{+    0.0013+    0.0026}$  & $    0.1203_{-    0.0015-    0.0029}^{+    0.0015+    0.0029}$  \\

$\Omega_b h^2$ & $    0.02240_{-    0.00015-    0.00028}^{+    0.00014+    0.00029}$ & $    0.02238_{-    0.00014-    0.00028}^{+    0.00014+    0.00028}$ & $    0.02238_{-    0.00015-    0.00029}^{+    0.00015+    0.00031}$  & $    0.02239_{-    0.00015-    0.00029}^{+    0.00015+    0.00030}$  \\

$100\theta_{MC}$ & $    1.04095_{-    0.00033-    0.00060}^{+    0.00032+    0.00061}$ & $    1.04096_{-    0.00032-    0.00061}^{+    0.00031+    0.00064}$  & $    1.04094_{-    0.00031-    0.00062}^{+    0.00033+    0.00060}$    & $    1.04090_{-    0.00030-    0.00059}^{+    0.00031+    0.00057}$  \\

$\tau$ & $    0.0530_{-    0.0073-    0.015}^{+    0.0073+    0.016}$ &  $    0.0546_{-    0.0075-    0.017}^{+    0.0077+    0.017}$ & $    0.0541_{-    0.0077-    0.016}^{+    0.0077+    0.017}$   & $    0.0544_{-    0.0086-    0.016}^{+    0.0075+    0.016}$  \\

$n_s$ & $    0.9655_{-    0.0043-    0.0088}^{+    0.0046+    0.0084}$ & $    0.9659_{-    0.0040-    0.0087}^{+    0.0045+    0.0081}$ & $    0.9651_{-    0.0042-    0.0084}^{+    0.0042+    0.0083}$   & $    0.9644_{-    0.0048-    0.0086}^{+    0.0044+    0.0089}$ \\

${\rm{ln}}(10^{10} A_s)$ & $    3.042_{-    0.015-    0.033}^{+    0.015+    0.033}$ &   $    3.045_{-    0.016-    0.034}^{+    0.016+    0.035}$  & $    3.044_{-    0.016-    0.033}^{+    0.016+    0.033}$  & $    3.046_{-    0.017-    0.029}^{+    0.015+    0.032}$  \\

$\Delta$ & $>3.0\,>0.33$ & $    0.26_{-    0.40-    0.73}^{+    0.37+    0.79}$ & $    1.56_{-    0.38-    0.67}^{+    0.35+    0.72}$   & $    0.85_{-    0.41-    0.79}^{+    0.44+    0.71}$   \\

$\Omega_{m0}$ & $    0.207_{-    0.061-    0.07}^{+    0.021+    0.10}$ &  $    0.304_{-    0.012-    0.023}^{+    0.012+    0.024}$  & $    0.260_{-    0.011-    0.020}^{+    0.010+    0.020}$   & $    0.285_{-    0.0098-    0.018}^{+    0.0093+    0.019}$   \\

$\sigma_8$ & $    0.958_{-    0.043-    0.148}^{+    0.098+    0.110}$ & $    0.822_{-    0.021-    0.037}^{+    0.018+    0.038}$  & $    0.876_{-    0.017-    0.034}^{+    0.017+    0.033}$  & $    0.848_{-    0.020-    0.039}^{+    0.023+    0.036}$  \\

$H_0$ & $   85_{-    6-   17}^{+   12+   13}$ & $   68.6_{-    1.5-    2.7}^{+    1.3+    2.9}$  & $   74.3_{-    1.4-    2.7}^{+    1.4+    2.8}$  & $   71.0_{-    1.3-    2.6}^{+    1.4+    2.5}$  \\

$S_8$ & $    0.785_{-    0.035-    0.050}^{+    0.022+    0.058}$ &  $    0.827_{-    0.014-    0.026}^{+    0.013+    0.028}$  & $    0.814_{-    0.015-    0.029}^{+    0.015+    0.030}$  & $    0.826_{-    0.014-    0.026}^{+    0.014+    0.027}$  \\

$r_{\rm{drag}}$ & $ 147.10_{-    0.30-    0.58}^{+    0.30+    0.58}$ &  $  147.13_{-    0.29-    0.57}^{+    0.30+    0.57}$  & $  147.06_{-    0.29-    0.56}^{+    0.29+    0.59}$  & $  146.99_{-    0.33-    0.64}^{+    0.32+    0.64}$   \\

\hline\hline                                                                                                                    
\end{tabular}                                                                                                                   
\caption{Observational constraints at 68\% and 95\% CL on the free and derived parameters of the GEDE scenario considering CMB from Planck 2018, and its combination with BAO and R19. }
\label{tab:GEDEI}                                                                                                   
\end{table*}                                                                                                                     
\end{center}                                                                                                                    
\endgroup
\begingroup                                                                                                                     
\squeezetable                                                                                                                   
\begin{center}                                                                                                                  
\begin{table*}                                                                                                                   
\begin{tabular}{ccccccccc}                                                                                                            
\hline\hline                                                                                                                    
Parameters & Planck 2018+JLA & Planck 2018+Pantheon & Planck 2018+BAO+JLA+R19 & Planck 2018+BAO+Pantheon+R19\\ \hline

$\Omega_c h^2$ & $    0.1202_{-    0.0013-    0.0026}^{+    0.0013+    0.0025}$ & $    0.1203_{-    0.0014-    0.0026}^{+    0.0014+    0.0027}$ & $    0.1201_{-    0.0012-    0.0024}^{+    0.0012+    0.0023}$ & $    0.1199_{-    0.0012-    0.0024}^{+    0.0012+    0.0023}$  \\

$\Omega_b h^2$ & $    0.02235_{-    0.00014-    0.00029}^{+    0.00015+    0.00029}$ & $    0.02236_{-    0.00015-    0.00029}^{+    0.00015+    0.00030}$ & $    0.02238_{-    0.00016-    0.00028}^{+    0.00014+    0.00030}$ & $    0.02240_{-    0.00015-    0.00029}^{+    0.00014+    0.00029}$  \\

$100\theta_{MC}$ & $    1.04090_{-    0.00031-    0.00062}^{+    0.00031+    0.00062}$ & $    1.04092_{-    0.00031-    0.00059}^{+    0.00031+    0.00061}$ & $    1.04096_{-    0.00030-    0.00060}^{+    0.00031+    0.00060}$ & $    1.04095_{-    0.00031-    0.00061}^{+    0.00031+    0.00058}$  \\

$\tau$ & $    0.0544_{-    0.0074-    0.014}^{+    0.0075+    0.015}$ & $    0.0541_{-    0.0075-    0.015}^{+    0.0074+    0.015}$ & $    0.0543_{-    0.0078-    0.016}^{+    0.0074+    0.015}$  & $    0.0550_{-    0.0077-    0.015}^{+    0.0076+    0.016}$  \\

$n_s$ & $    0.9647_{-    0.0047-    0.0084}^{+    0.0043+    0.0085}$ & $    0.9647_{-    0.0043-    0.0089}^{+    0.0047+    0.0086}$ & $    0.9649_{-    0.0041-    0.0079}^{+    0.0041+    0.0081}$  & $    0.9656_{-    0.0041-    0.0080}^{+    0.0041+    0.0084}$  \\

${\rm{ln}}(10^{10} A_s)$ & $    3.045_{-    0.016-    0.029}^{+    0.015+    0.031}$ & $    3.045_{-    0.016-    0.031}^{+    0.016+    0.031}$  & $    3.044_{-    0.016-    0.031}^{+    0.015+    0.032}$  & $    3.046_{-    0.016-    0.030}^{+    0.016+    0.032}$  \\

$\Delta$ & $    0.30_{-    0.40-    0.73}^{+    0.36+    0.79}$ & $    0.25_{-    0.26-    0.52}^{+    0.26+    0.51}$  & $    0.69_{-    0.25-    0.48}^{+    0.25+    0.49}$  & $    0.55_{-    0.21-    0.43}^{+    0.20+    0.42}$   \\

$\Omega_{m0}$ & $    0.305_{-    0.015-    0.029}^{+    0.015+    0.029}$ & $0.307_{-    0.011-    0.021}^{+    0.011+    0.021}$  & $    0.289_{-    0.0072-    0.014}^{+    0.0074+    0.015}$  & $    0.293_{-    0.0067-    0.013}^{+    0.0065+    0.013}$  \\

$\sigma_8$ & $    0.825_{-    0.019-    0.034}^{+    0.017+    0.037}$ &  $    0.823_{-    0.013-    0.028}^{+    0.016+    0.028}$  & $    0.841_{-    0.014-    0.027}^{+    0.014+    0.027}$  & $    0.834_{-    0.012-    0.025}^{+    0.012+    0.025}$  \\

$H_0$ & $   68.6_{-    1.8-    3.0}^{+    1.5+    3.3}$ & $   68.3_{-    1.1-    2.1}^{+    1.1+    2.3}$  & $   70.38_{-    0.89-    1.7}^{+    0.87+    1.8}$  & $   69.86_{-    0.74-    1.4}^{+    0.75+    1.4}$  \\

$S_8$ & $    0.831_{-    0.015-    0.030}^{+    0.015+    0.031}$ & $    0.832_{-    0.015-    0.029}^{+    0.015+    0.030}$ & $    0.825_{-    0.013-    0.025}^{+    0.013+    0.026}$   & $    0.824_{-    0.013-    0.025}^{+    0.014+    0.025}$  \\

$r_{\rm{drag}}$ & $  147.05_{-    0.31-    0.55}^{+    0.28+    0.58}$ & $  147.03_{-    0.30-    0.57}^{+    0.30+    0.58}$  & $  147.05_{-    0.28-    0.54}^{+    0.27+    0.53}$  & $  147.10_{-    0.27-    0.54}^{+    0.27+    0.52}$  \\
\hline\hline                                                                                                                    
\end{tabular}                                                   \caption{Here we present the constraints on the same GEDE scenario but focusing on the effects of JLA and Pantheon on the model parameters. }\label{tab:GEDEII}                 
\end{table*}                                                
\end{center}                                                   
\endgroup     
\begin{figure*}
\centering
\includegraphics[width=0.6\textwidth]{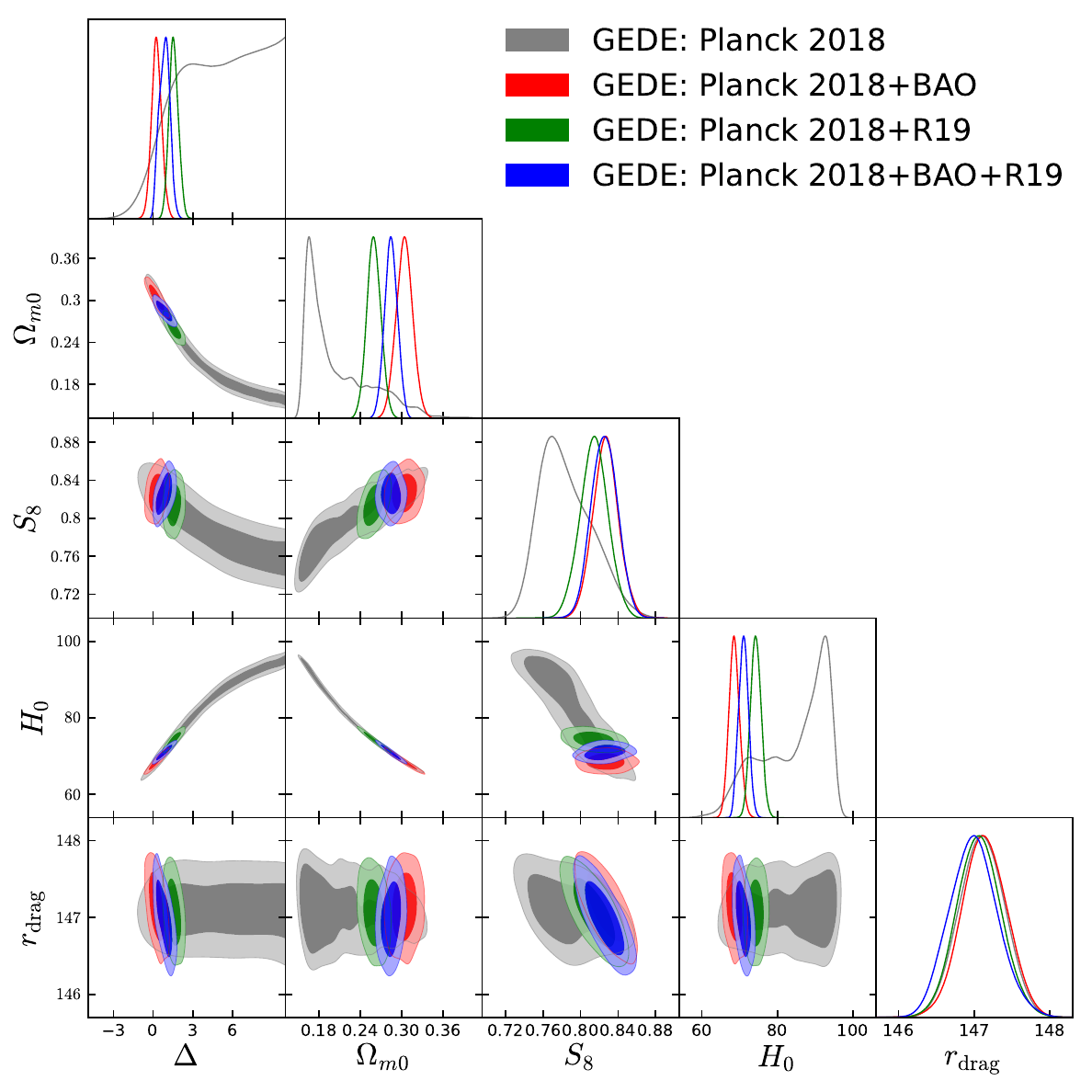}
\caption{We show the triangular plot containing 1-dimensional marginalized posterior distributions and the 2-dimensional joint contours for some important parameters of the GEDE scenario considering the datasets Planck 2018, Planck 2018+BAO, Planck 2018+R19 and Planck 2018+BAO+R19. }
\label{fig-gede}
\end{figure*}
\begin{figure*}
\centering
\includegraphics[width=0.6\textwidth]{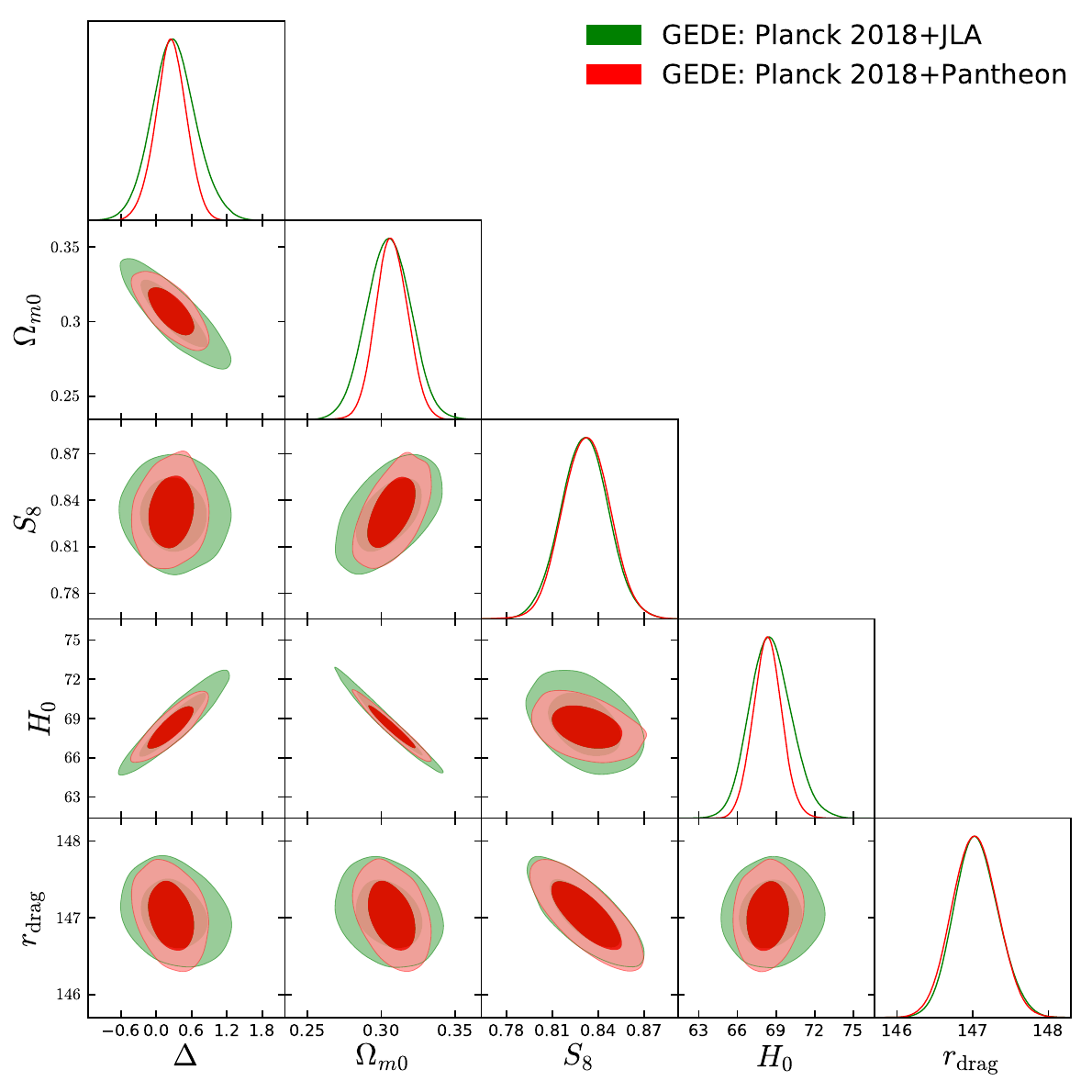}
\caption{Here we show the triangular plot for the GEDE scenario for Planck 2018+Pantheon and Planck 2018+JLA datasets in order to predict mainly the individual effects of Pantheon and JLA. }
\label{fig-gede-snia}
\end{figure*}

\section{Observational Results}
\label{sec-results}

Let us summarize the main results of the GEDE model that we find using a variety of cosmological probes. 
In Tables \ref{tab:GEDEI} and \ref{tab:GEDEII}, we have shown the 68\% and 95\% CL constraints on the free and derived parameters of the GEDE scenario using the CMB spectra from Planck 2018 and its combination with other external cosmological probes. 

In the second column in Table \ref{tab:GEDEI}, we show the results from Planck 2018 alone. We find that the Hubble constant assumes a much larger mean value than R19, but with relaxed error bars, thus, solving the  Hubble tension within $2\sigma$. Because of the anti-correlation between $H_0$ and $\Omega_{m0}$, $\Omega_{m0}$ assumes a very small value ($\Omega_{m0} = 0.207_{-0.061}^{+    0.021}$ at 68\% CL). The only free parameter of this model, $\Delta$, has an upper bound ($\Delta>3$ at 68\% CL and $\Delta>0.33$ at 95\% CL) and this upper bound implies that the $\Lambda$CDM model (for which $\Delta = 0$) is excluded at more than 95\% CL, while the PEDE model (for which $\Delta = 1$) is excluded at more than 68\% CL. Thus, as we see, Planck 2018 data rule out the $\Lambda$CDM scenario.   
If we compare this full Planck 2018 analysis with the previous work \cite{Li:2020ybr} without perturbations, we can see that the $H_0$ constraint changes significantly, and now it is more in agreement with R19. The very same thing happens for $\Delta$, highlighting the importance of a robust data analysis. This means that we are playing here with the $\Delta-H_0$ positive degeneracy (see Fig.~\ref{fig-gede}), and introducing perturbations, we allow for larger values of both $\Delta$ and $H_0$ at the same time, as well as a lower value of $S_8$ solving also the tension with the cosmic shear data.

When the BAO data are added to the CMB from Planck 2018 (see the third column in Table \ref{tab:GEDEI}), $H_0$ goes down and the existing tension with R19 is now reduced to $2.8 \sigma$. Recall that for Planck 2018 alone, the tension of $H_0$ with R19 was alleviated within $2 \sigma$. 
The free parameter $\Delta$ is well constrained and the scenario now becomes consistent with the $\Lambda$CDM model within $1\sigma$ ($\Delta  = 0.26_{-    0.40}^{+    0.37}$ at 68\% CL) and with PEDE within $2\sigma$.   

For Planck 2018+R19, $\Delta \neq 0$ at more than 95\% CL, that means, $\Lambda$CDM is ruled out at more than two standard deviations. However, we are now in agreement with $\Delta=1$, the PEDE scenario, within $2\sigma$. Interestingly, when we consider the full Planck 2018+BAO+R19 combination, shown in the last column in Table \ref{tab:GEDEI}, not only is $\Lambda$CDM ruled out at more than 95\% CL, but this model is in agreement within $1\sigma$ with $\Delta=1$, this means that PEDE is the favored model by this combination of data.

At this point, we can discuss the impacts on the constraints when JLA and Pantheon datasets are separately added to the CMB. As expected we have similar results for these combinations (see Fig.~\ref{fig-gede-snia}), with only marginal differences, and these are shown in Table \ref{tab:GEDEII}. For both Planck 2018+JLA and Planck 2018+Pantheon (see Fig. \ref{fig-gede-snia}), $\Delta = 0$ is consistent within $1\sigma$, and thus recovering the $\Lambda$CDM model, but while Planck 2018+JLA is in agreement with PEDE $\Delta=1$ within  95\% CL, Planck 2018+Pantheon is ruling out this model at more than $2\sigma$. Similar considerations hold for the Hubble constant tension: Both the datasets combination give a similar mean value for $H_0$, but Planck 2018+Pantheon has stronger error bars. This translates into a $2.6\sigma$ tension for Planck 2018+JLA, and a $3.2\sigma$ tension for Planck 2018+Pantheon with respect to R19.

For the two full dataset combinations, namely, Planck 2018+BAO+JLA+R19 and Planck 2018+BAO+Pantheon+R19, $\Delta$ is found to be different from zero at more than 95\% CL. This result is mainly driven by the R19 dataset as one can quickly realize by looking at various datasets in Tables \ref{tab:GEDEI} and \ref{tab:GEDEII}. 
However, as for the previous cases,
Planck 2018+BAO+JLA+R19 is in agreement with PEDE (corresponds to $\Delta=1$) within 95\% CL, but Planck 2018+BAO+Pantheon+R19 is ruling out PEDE too at more than $2\sigma$. 

Finally, we compute the Bayesian evidence summarizing the results in Table~\ref{tab:BE}, to quantify which model performs better in the fit of the data between $\Lambda$CDM and GEDE.
We use the publicly available cosmological package \texttt{MCEvidence}\footnote{\href{https://github.com/yabebalFantaye/MCEvidence}{github.com/yabebalFantaye/MCEvidence}~\cite{Heavens:2017hkr,Heavens:2017afc}.}, and compute the evidences in terms of the Bayes factor $\ln B_{ij}$ where 
its negative (positive) values indicate a preference of the data for the $\Lambda$CDM (GEDE) scenario. To interpret the numerical values of $\ln B_{ij}$, we use instead the revised Jeffreys scale by Kass and Raftery as in Refs.~\cite{Kass:1995loi,Trotta:2008qt}, for which if $0 \leq | \ln B_{ij}|  < 1$ the model has inconclusive evidence, if $1 \leq | \ln B_{ij}|  < 2.5$ the model has  weak evidence, if $2.5 \leq | \ln B_{ij}|  < 5$ the model has  moderate evidence, and if $| \ln B_{ij} | \geq 5$ the model has  strong evidence. Looking at Table~\ref{tab:BE} we can see that the GEDE model is mostly preferred by the fitting of the data for Planck 2018 alone and when the R19 prior is included with Planck 2018, while for the other combined analyses, namely Planck 2018+BAO, Planck 2018+JLA and Planck 2018+Pantheon, we have an inconclusive result.  

Even though the emergent DE model has not much effect on the early universe, there are still substantial differences when calculating the Integrated Sachs-Wolfe  effect and the low $\ell$ angular power spectrum in comparison with the cosmological constant~\cite{Li:2020ybr}. 
These differences might be the main reason that the GEDE model can provide a better fit to Planck data alone with respect to the standard $\Lambda$CDM model. More analyses on the GEDE model should be performed in order to gain a full understanding of this model. 

\begin{table}[h]
\centering
\begin{tabular}{|c|c|c|c|c|}
\hline 
Data & $\ln B_{ij}$\\
\hline 
Planck 2018       &       $2.9$ \\
Planck 2018+BAO     &       $0.8$ \\
Planck 2018+R19     &       $12.1$ \\
Planck 2018+BAO+R19   &     $7.9$ \\
Planck 2018+JLA         &   $-0.2$ \\
Planck 2018+Pantheon      & $-0.9$ \\
Planck 2018+BAO+JLA+R19  &  $6.1$ \\
Planck 2018+BAO+Pantheon+R19 &   $5.8$ \\

\hline 
\end{tabular}
\caption{The table shows the values of $\ln B_{ij}$ calculated for the GEDE model with respect to the $\Lambda$CDM as the reference model. The negative value in $\ln B_{ij}$ indicates that $\Lambda$CDM is preferred over GEDE while the positive sign denotes the opposite.}
\label{tab:BE}
\end{table}
\section{Concluding remarks}
\label{sec-conclu}

No doubt, $\Lambda$CDM has received notable attention over the years for its ability to fit most of the observational data, but the cosmological constant problem has still remained unanswered. Even if we forget the cosmological constant issue, some recent observations indicate that we cannot totally rely on the $\Lambda$CDM-driven cosmology -- the measurements of some key cosmological parameters, namely $H_0$ and $\sigma_8$, within this paradigm, do not match with their measurements made by other astronomical probes, and are many sigmas apart from one another. This is widely known as the cosmological tension. These discrepancies at many standard deviations cannot be avoided, and, if not due to systematic errors in the experiments, they actually hint for some limitations of the $\Lambda$CDM cosmology. Therefore, modifications of the $\Lambda$CDM scenario are welcomed, since there should not be any reason to favor it over others. If a new cosmological model can solve those tensions, or at least one of the tensions, then the model should be investigated robustly in order to understand how it fits overall with the expansion history of the Universe. However, it is equally true that the cosmological models beyond $\Lambda$CDM usually contain some extra free parameters than six (number of free parameters in a flat $\Lambda$CDM model), and for this reason, sometimes, the model comparison disfavors them in comparison to the standard $\Lambda$CDM model. In this context we recall PEDE \cite{Li:2019yem} having exactly six free parameters. The model can excellently solve the $H_0$ tension within $1\sigma$ \cite{Pan:2019hac}, and, for some combination of the cosmological probes, this model is preferred by the data over the $\Lambda$CDM model as well. Having such excellent features, an extension of this model, called generalized emergent DE model, was proposed and investigated in Ref.  \cite{Li:2020ybr}.   
The generalized emergent DE model introduces one extra free parameter, $\Delta$, compared to the PEDE model, which can discriminate between the $\Lambda$CDM model (recovered for $\Delta=0$) and the PEDE model (recovered for $\Delta=1$).
In this way we can determine which model is preferred by the fit of the data.

We have constrained the present GEDE scenario considering its  evolution at the level of linear perturbations and with the use of latest cosmological probes from Planck 2018, BAO, SNIa (both JLA and Pantheon samples) and R19. The results are summarized in Tables \ref{tab:GEDEI} and \ref{tab:GEDEII}. 
We find that, while the $\Lambda$CDM model is ruled out at more than 95\% CL for most of the cases, with the exception of Planck 2018+BAO or Planck 2018+SNIa, PEDE is in agreement with the data for Planck 2018+BAO+R19 combination within $1\sigma$, and with Planck + R19, Planck 2018+JLA, and Planck 2018+JLA+BAO+R19 within $2\sigma$.

The model has also been assessed through the Bayesian evidence analysis which reports a preference of the GEDE model over the $\Lambda$CDM by most of the observational datasets but not for all (see Table~\ref{tab:BE}). For instance, the inclusion of JLA or Pantheon or BAO to Planck 2018 gives an inconclusive result. 
However, based on the present results, it is clear that the GEDE model is worth investigating.  
Therefore, the present work clearly tells us that the cosmological models having the ability to reconcile the $H_0$ tension with a very good fitting of the data are quite appealing even if they are motivated from the phenomenological ground.  Last but not least, we must say that there is no reason to exclude the phenomenology for two reasons: First, the DE sector is totally unknown, and second, since we have a lot of observational data, we have enough space to judge the viability of the model arising from the phenomenological ground. A model can be thrown at any time, but before throwing it, one needs to replace it with something good. Until we win this game, we have to be open minded.    

\acknowledgments
The authors thank the referee for raising some important queries which improved the article. 
WY was supported by the National Natural Science Foundation of China under Grants No. 12175096 and No. 11705079, and Liaoning Revitalization Talents Program under Grant no. XLYC1907098.
EDV acknowledges the support of the Addison-Wheeler Fellowship awarded by the Institute of Advanced Study at Durham University. SP gratefully acknowledges the Mathematical Research Impact-Centric Support Scheme (File No. MTR/2018/000940) funded by the Science and Engineering Research Board, Govt. of India,.
AS would like to acknowledge the support of the Korea Institute for Advanced Study (KIAS) grant funded by the Korea government.
X. Li was supported by National Natural Science Foundation of China under Grants Nos. 12003006, Hebei NSF under Grant No. A2020205002.
\bibliography{biblio}
\end{document}